\documentclass[preprint,3p]{elsarticle}
\usepackage{epstopdf}
\usepackage{caption}
\usepackage{setspace}
\usepackage{amssymb}
\usepackage{multirow}
\usepackage{amsthm}
\usepackage{booktabs}
\usepackage{amsmath}
\usepackage{amssymb}
\usepackage{mathrsfs}
\usepackage{longtable}
\usepackage{lscape}
\usepackage{url}
\usepackage{tabularx}
\usepackage{epsfig}
\usepackage{colortbl}
\usepackage{subfigure}
\usepackage{tikz,mathpazo}
\usepackage{graphicx}
\usepackage{natbib}
\usepackage[noend]{algpseudocode}
\usepackage{algorithm}
\usepackage{algorithmicx}
\usepackage{hyperref}
\usepackage [latin1]{inputenc}

\newtheorem{definition}{Definition}[section]
\usetikzlibrary{shapes.geometric, arrows}
\biboptions{numbers,sort&compress}
\journal{}

\begin{document}
\begin{frontmatter}
\title{Evaluating importance of nodes in complex networks with local volume information dimension}
\author[address1]{Hanwen Li}
\author[address1,address5]{Qiuyan Shang}
\author[address1]{Fangzheng Duan}
\author[address1,address2,address3,address4]{Yong Deng \corref{label1}}

\address[address1]{Institute of Fundamental and Frontier Science, University of Electronic Science and Technology of China, Chengdu 610054, China}
\address[address5]{Yingcai Honors of school, University of Electronic Science and Technology of China, Chengdu, 610054, China}
\address[address2]{School of Education, Shaanxi Normal University, Xi'an, 710062, China}
\address[address3]{School of Knowledge Science, Japan Advanced Institute of Science and Technology, Nomi, Ishikawa 923-1211, Japan}
\address[address4]{Department of Management, Technology, and Economics, ETH Zrich, Zurich, Switzerland}

\cortext[label1]{Corresponding author at: Institute of Fundamental and Frontier Science, University of
	Electronic Science and Technology of China, Chengdu, 610054, China. E-mail: dengentropy@uestc.edu.cn,
	prof.deng@hotmail.com.(Yong Deng)}
\begin{abstract}
How to evaluate the importance of nodes is essential in research of complex network. There are many methods proposed for solving this problem, but they still have room to be improved. In this paper, a new approach called local volume information dimension is proposed. In this method, the sum of degree of nodes within different distances of central node is calculated. The information within the certain distance is described by the information entropy. Compared to other methods, the proposed method considers the information of the nodes from different distances more comprehensively. For the purpose of showing the effectiveness of the proposed method, experiments on real-world networks are implemented. Promising results indicate the effectiveness of the proposed method.
\end{abstract}
\begin{keyword}
Complex network, Node importance, Local volume information dimension
\end{keyword}
\end{frontmatter}
\section{Introduction}

In recent days, complex network theory is deeply studied by massive researchers. Complex network theory can be used in many practical fields like economics \cite{gao2019computational}, biology \cite{liu2020computational,cheong2019paradoxical}, chemistry \cite{chrayteh2021disentangling}, sociology \cite{lai2020social} effectively. There are many challenging topics in complex networks like link prediction \cite{berahmand2021modified}, scale free property \cite{rak2020fractional}, fractal dimension \cite{wen2021fractal,Gao2021Information}, game theory \cite{zhu2020role,shen2021exit,cheong2020relieving} and so forth. What is more, complex theory can be used to solve other types of topics or problems, including time series analysis \cite{liu2020fuzzy,unknown}, uncertainty modeling \cite{zhao2020complex,xiong2021conflicting,chen2021probability}. We are witnessing and looking forward to the rapid development of complex networks \cite{vega2020influence}.

In complex network, different nodes have different importance. Important node should have stronger effect on the network. Finding important nodes is useful in many areas, like disease control \cite{doostmohammadian2020centrality}, industrial engineering \cite{li2021research} and so on. Hence, it is necessary to evaluate the importance of nodes correctly. There are many methods proposed for evaluating importance of nodes. Li and Deng \cite{li2021local} propose loval volume dimension based on power law theory and fractal theory. Li et al. \cite{li2019identifying} get inspiration from gravitivation law and propose gravity model which considers both global and local information of nodes. a generalized mechanics model, which is proposed by Liu et al. \cite{liu2020gmm}, is also named weighted gravity model. Weighted gravity model considers both the self importance and the influence from other nodes. Li et al. \cite{li2021generalized} integrates local clustering coefficient and gravity model and propose generalized gravity model subsequently. Generalized gravity model is also the generalized form of gravity model, which is more flexible. Li and Xiao \cite{li2021identification} propose effective gravity model based on information entropy. Density centrality is proposed by Ibnoulouafi et al. \cite{ibnoulouafi2018density} based on density formula. Berahmand et al. \cite{berahmand2018new} explore both the positive and negative effect of the local clustering coefficient in evaluating node importance. Zareie and Sheikhahmadi \cite{zareie2018hierarchical} develop a hierarchical approach to mine vital nodes. Wang et al. \cite{wang2021identifying} use discrete Moth-Flame optimization algorithm to find vital nodes. Tang et al. \cite{tang2020research} propose k-order propogation number algorithm to investigate vital spreaders.

The main contribution of this paper is to propose a new approach called local volume information dimension (LVID) to evaluate importance of nodes. In this method, the sum of nodes' degree within different distances of central node is considered. The information within the certain distance is drescribed by the Shannon entropy \cite{shannon1948mathematical}. The distribution of information is considered more comprehensively with the proposed method. The superiority of the local volume information dimension is solidated by experiments in the comparison with five other existing methods on real-world networks. 

The remaining part of this paper is introduced as follows. In Section 2, preliminaries in this paper are briefly introduced. Proposed method is introduced in Section 3. Experiments on real-world networks are used to verify whether the proposed method works in Section 4. Finally, conclusions will be made in Section 5.
\section{Preliminaries}

In this section, some preliminaries of this paper including complex network theory, local volume dimension, susceptible-infectious-recovered (SIR) model, Kendall correlation coefficient will be introduced.

\subsection{Complex network theory}
Complex network is represented in the form of a graph. A complex network is a graph $G = (V, E)$, where $V$ denotes the vetices of the graph and $E$ are the edges of the graph. $A$ is the adjacency matrix of graph. If node $i$ and node $j$ are connected, $A_{ij} = 1$ otherwise 0. $d_{ij}$ refers to the distance between node $i$ and node $j$ and can be calculated by the shortest path algorithms. $d_{ij}$ are different for different nodes.

\subsection{Kendall correlation coefficient}
Kendall correlation coefficient \cite{kendall1945treatment}, which is also called $\tau$ value, is an efficient methmatical measure to measure the correlation betweeen two series. It is defined as follows.

\begin{definition}
	Given two series $X$ and $Y$ with the same length. $X = \left(x_1, x_2, ..., x_n\right)$ and $Y = \left(y_1, y_2, ..., y_n\right)$, the pair $(x_i, x_j)$ and $(y_i, y_j)$ is called consistent if $x_i < x_j$ and $y_i < y_j$ or $x_i > x_j$ and $y_i > y_j$ otherwise inconsistent. The Kendall correlation coefficient can be cauculated as follows.
	\begin{equation}
	tau(X,Y)=\frac{2(n_p-n_n)}{n(n-1)}
	\end{equation}
	where $n_p$ is the number of consistent pairs and $n_n$ is the number of inconsistent pairs. $n$ is the size of the series. Higher value of Kendall correlation coefficient refers to higher correlation between two series.
\end{definition}

\subsection{SIR model}
Susceptible-infected recovered (SIR) model is used to simulate the whole process of disease propagation on the network \cite{hethcote2000mathematics}. The process of SIR model is illustrated in Figure 1. Each node in SIR model has three states, susceptible state, infected state and recovered state. Nodes in susceptible state can be infected by other infected nodes. Infected nodes can be recovered and recovered nots can not be infected again. There are some parameters of this model, the simulation time $T$ means the number of simulation steps of the SIR model. The infection rate $\beta$ is the probability of the susceptible node infected by adjacent infected nodes. The recovered rate $\gamma$ denotes the probability of infected nodes to recover. When $\gamma = 0$, the SIR model degenerates into susceptible infected (SI) \cite{zhou2006behaviors} model. $N$ is the number of independent experiments.

\begin{figure}
	\centering

	\subfigure{\includegraphics[]{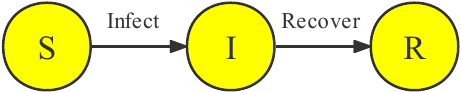}}
	\caption{The process of SIR model}
\end{figure}

\subsection{Local volume dimension \cite{li2021local}}
Local volume dimension \cite{li2021local}, which is proposed by Li and Deng, is an effective approach for node importance evaluation. It is defined as follows.

\begin{equation}
LVD(i) = -\frac{d}{dlnl}ln s_i(l)
\end{equation}

where $s_i(l)$ is the sum of degree of nodes within the distance $l$ from the node $i$. $l$ ranges from 1 to the maximum value of the distances from node $i$ and other nodes. 

\section{Proposed method}
In this section, the proposed method will be introduced. The proposed approach is defined as follows.

\begin{equation}
LVID(i) = -\frac{d}{dlnl}-\left(\frac{s_i(l)}{S}ln{\frac{s_i(l)}{S}}\right)
\end{equation}
where $s_i(l)$ is the sum of degree of nodes within the distance $l$ from the node $i$ and $S$ is the sum of the degrees of the network. $l$ ranges from 1 to the maximum value of the distances from node $i$ and other nodes. For calculation, LVID can be obtained from the slope of linear regression equation with $ln(l)$ and the entropy measure.

The proposed method considers information of the central node from different topological distances, and information from different distances is measured by information entropy. Information from different distances' nodes is considered. As we can see, the differences between LVID and LVD is that LVID consider the information in the certain distance by the information entropy. The information entropy can better reflect the distribution of information within different distances. Hence, LVID considers the information of different distances more comprehensively than LVD.

\section{Experiments}
To verify the superiority of the proposed method, experiments on six real-world networks are carried out. The six networks are Jazz network (198 nodes, 2742 edges), NS network  (379 nodes, 914 edges), PB network  (1222 nodes, 16714
edges), Celegans network (297 nodes, 2359 edges), Infectious network  (410 nodes, 17298 edges), PDZbase network (212 nodes, 2672 edges) respectively. LVID is compared with five other methods, including local volume dimension (LVD) \cite{li2021local}, generalized gravity model (GG) \cite{li2021generalized}, weighted gravity model (WG) \cite{liu2020gmm}, betweenness centrality (BC) \cite{newman2005measure} and pagerank algorithm (PR) \cite{brin1998anatomy}. 

\begin{figure}
	\centering
	\subfigure{\includegraphics[scale=0.385]{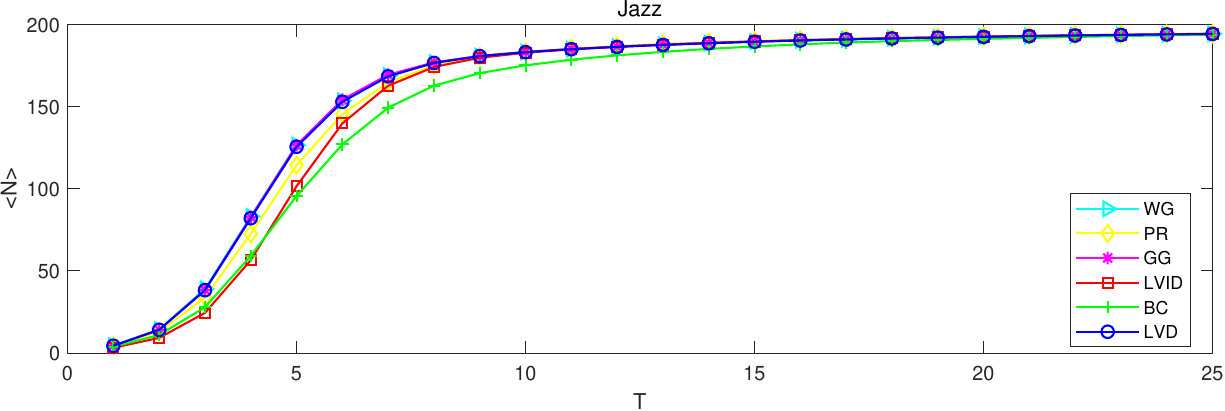}}
	\subfigure{\includegraphics[scale=0.385]{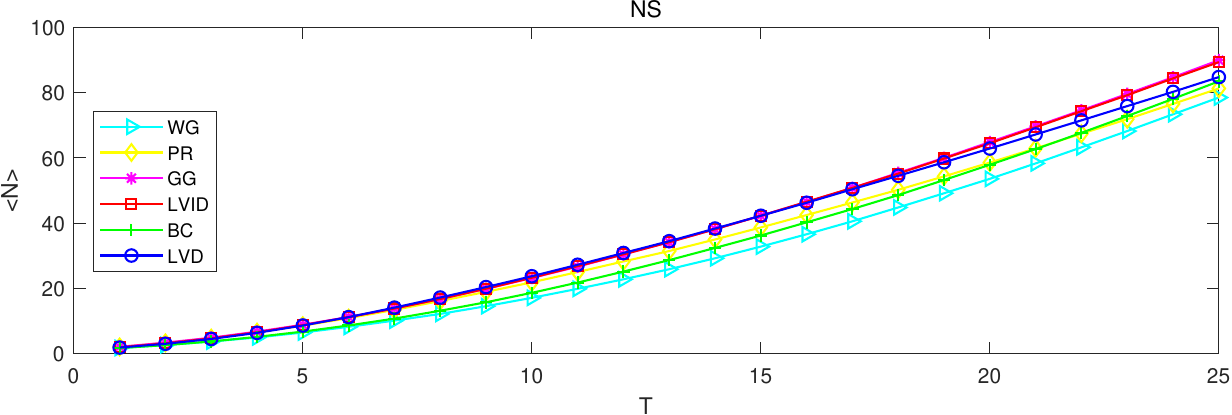}}
	\subfigure{\includegraphics[scale=0.385]{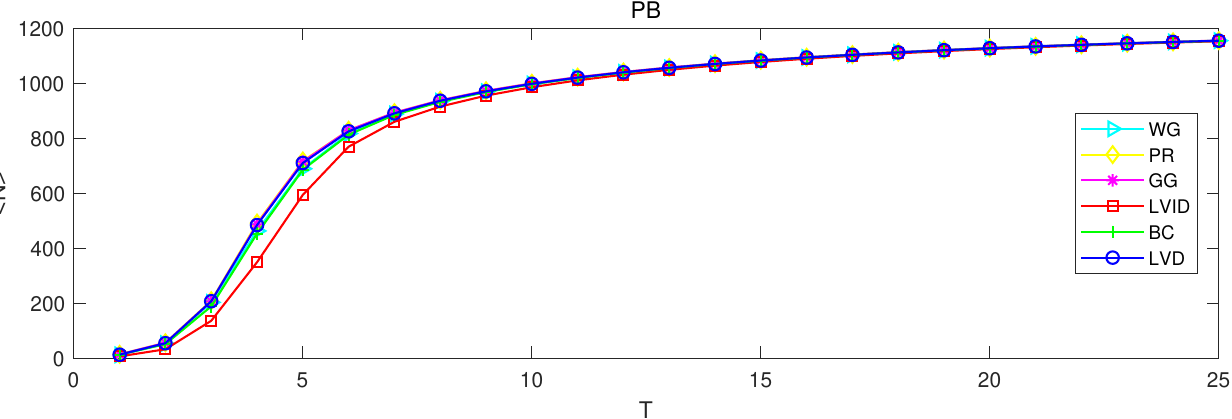}}
	\subfigure{\includegraphics[scale=0.385]{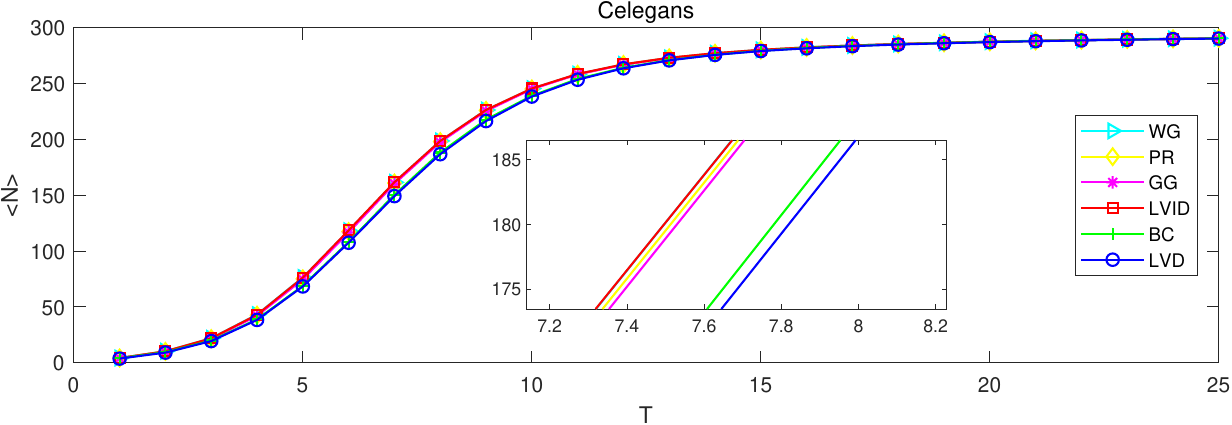}}
	\subfigure{\includegraphics[scale=0.385]{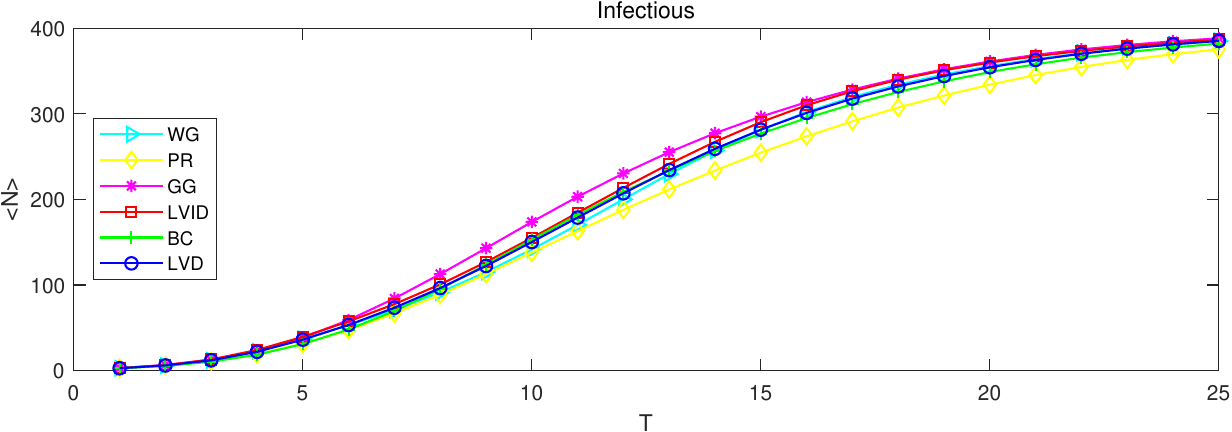}}
	\subfigure{\includegraphics[scale=0.385]{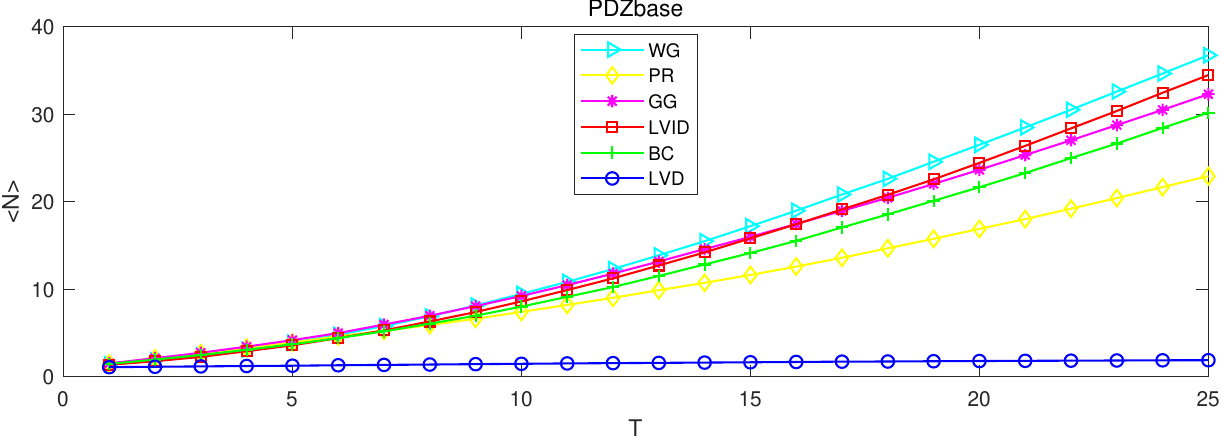}}
	\caption{Spreading abilities of different centrality measures in networks}
\end{figure}

\subsection{The infection ability of top-10 ranked nodes}
The infection ability of centrality measures are modeled by SIR model. The times of independent experiments is set as 100. The infection rate $\beta$ is set with the value of 0.05 and the recover rate $\gamma$ is 0 in this experiment. The number of steps for the SIR model $T$ equals to 25 as the initial process of disease propagating is more influential at most times. After $T$ steps, if the number of infected nodes is more, the centrality measure should have better ability to evaluate importance of nodes. The results are shown in Figure 2. 

In Jazz network, LVD and WG perfoms the best and LVD performs better than BC. In NS network, GG and LVID performs better than other methods by a large margin. In PB network, LVD, GG and PR have good performance. In Celegans network, LVID beats all other methods and reaches the best performance. In Infectious network, GG performs best and LVID outperforms all other methods. In PDZbase network, the proposed LVID method is only beaten by WG. Generally speaking, LVID has good performance in most networks, which shows the effectiveness of LVID.

\subsection{The correlation with SIR model}
In this experiment, compare the correlation the centrality measures with the spreading abilility of nodes in SIR model. If the correlation is higher, the centrality measure should have better performance. In this experiment, the time step $T$ is set to be 25. The experiment is executed for 100 times independently in order to reduce the effect of isolated results.  Kendell correlation coefficients of LVD and LVID between spreading abilitity of nodes in SIR model are compared.

\begin{figure}
	\centering
	\subfigure{\includegraphics[scale=0.6]{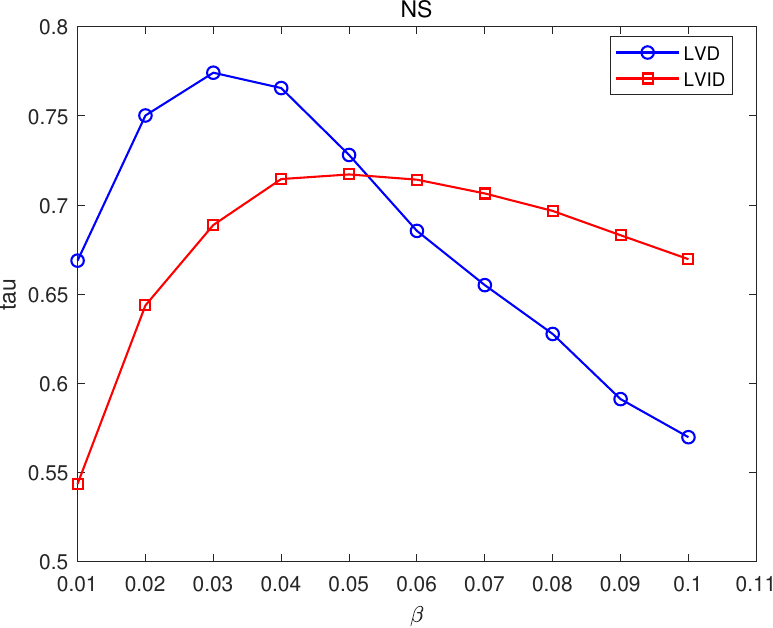}}
	\subfigure{\includegraphics[scale=0.6]{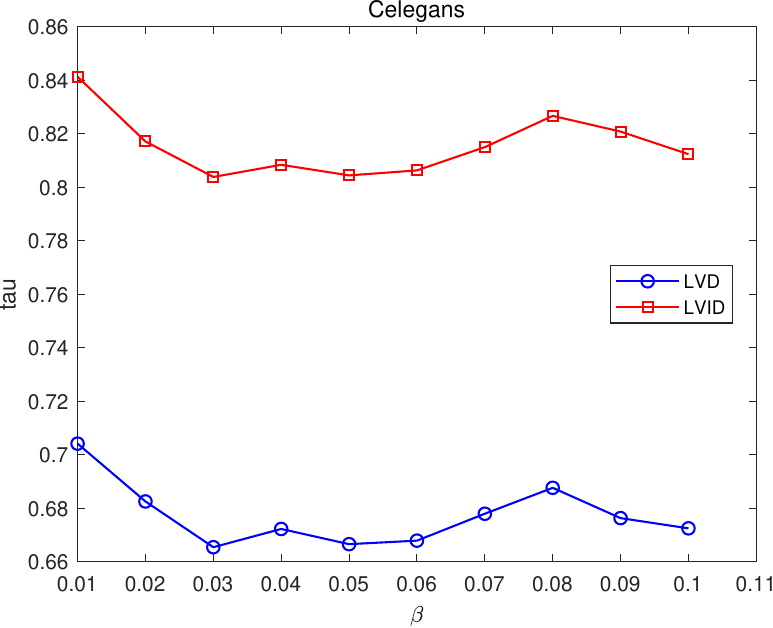}}
	\subfigure{\includegraphics[scale=0.6]{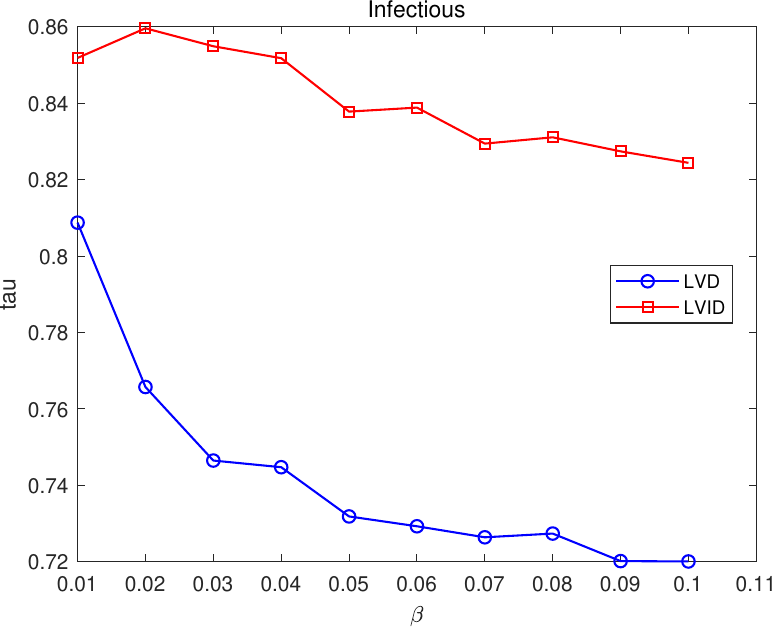}}
	\subfigure{\includegraphics[scale=0.6]{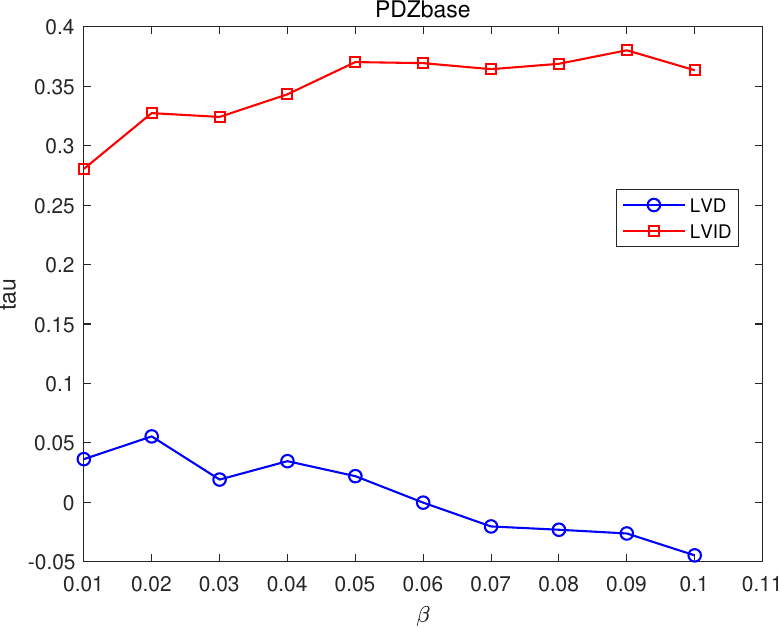}}
	\caption{The correlation of the centrality measures and spreading ability of nodes in SIR model with $\beta$ changes}
\end{figure}

First, we change the value of $\gamma$ and see the effect of $\beta$. $\gamma$ is set to be 0 and $\beta$ ranges from 0.01 to 0.1, the step length is 0.01. Results are shown in Figure 3 and Table 1-4. In NS network, when $\beta > 0.06$, LVID out performs LVD by a large margin. In Celegans network, LVID is always better than LVD. In Infectious network, LVID outperforms LVD in every time step. In PDZbase network, the performance of LVD is not competible with LVD. In a nutshell, LVID performs better than LVD, which shows the superiority of LVID.

Then, we fix $\beta$ as 0.05 and change the value of $\gamma$. $\gamma$ ranges from 0 to 1 and the step length is 0.1. Results are shown in Figure 4 and Table 5-8. Except the cases that $\gamma=0$ and $gamma=1$, LVD outperforms LVID. In other cases, LVID reaches better performance. In Celegans network, LVID is greater than LVD in every cases. In Infectious network, LVID putperforms LVD. In PDZbase network, LVID is better than LVD every time. We can conclude that LVID is more effective alternative compared with LVD in this experiment. 
\begin{figure}
	\centering
	\subfigure{\includegraphics[scale=0.6]{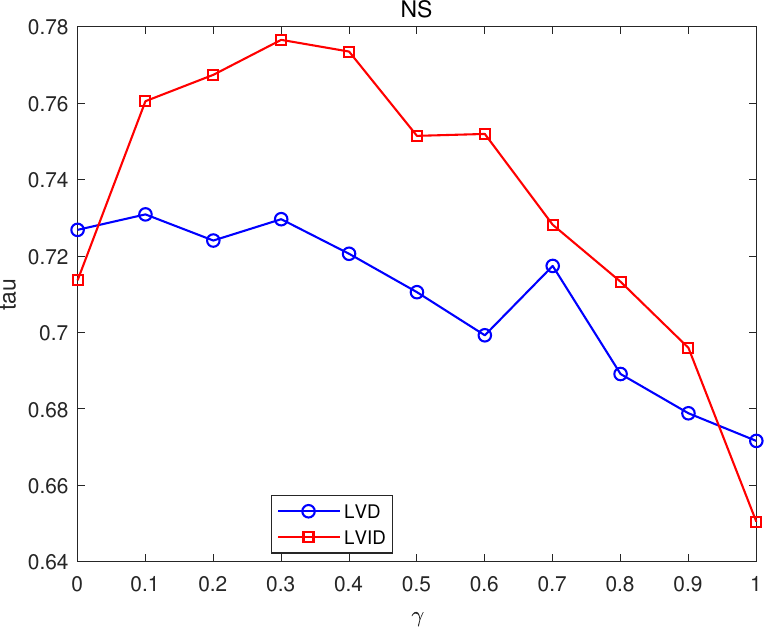}}
	\subfigure{\includegraphics[scale=0.6]{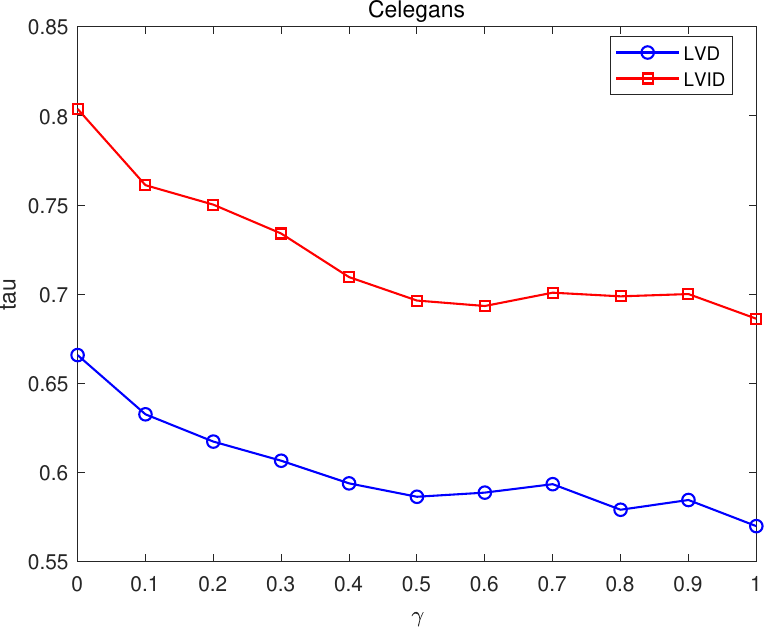}}
	\subfigure{\includegraphics[scale=0.6]{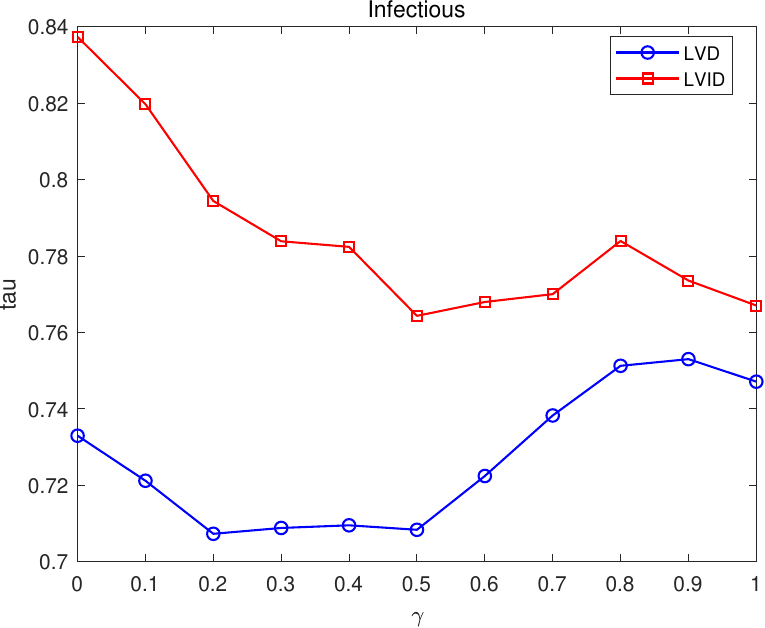}}
	\subfigure{\includegraphics[scale=0.6]{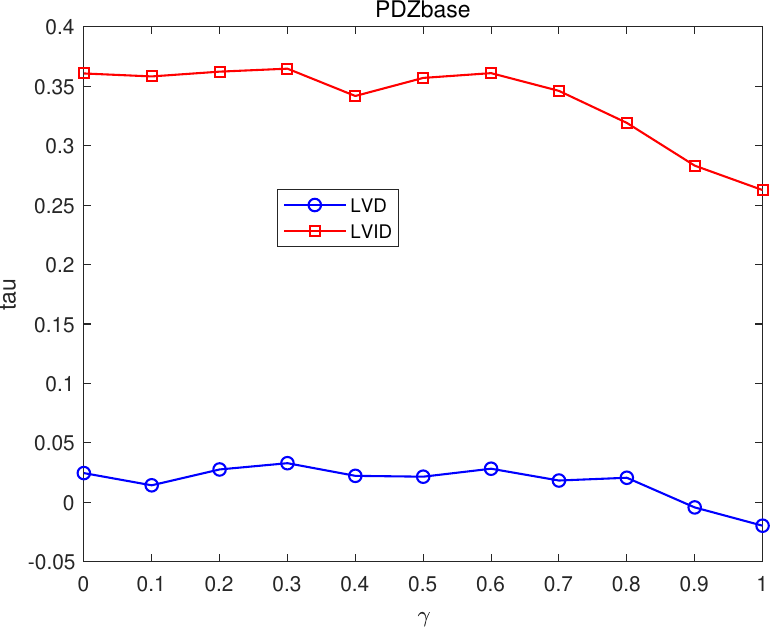}}
	\caption{The correlation of the centrality measures and spreading ability of nodes in SIR model with $\gamma$ changes}
\end{figure}

\begin{table*}[]
	\centering
	\caption{The Kendell correlation coefficient when $\beta$ changes in NS network}
	\begin{tabular}{ccccccccccc}
		\toprule
		 & 0.01 &0.02& 0.03& 0.04& 0.05&0.06&0.07&0.08&0.09&0.1 \\
		\midrule
 LVD \cite{li2021local}&\textbf{0.669}	&\textbf{0.750}	&\textbf{0.774}	&\textbf{0.766}	&\textbf{0.728}	&0.686	&0.655	&0.628	&0.591	&0.570 \\
 \midrule
  LVID &0.544	&0.644	&0.689	&0.715	&0.717	&\textbf{0.714}	&\textbf{0.707}	&\textbf{0.697}	&\textbf{0.683}	&\textbf{0.670} \\
		\bottomrule
	\end{tabular}
\end{table*}

\begin{table*}[]
	\centering
	\caption{The Kendell correlation coefficient when $\beta$ changes in Celegans network}
	\begin{tabular}{ccccccccccc}
		\toprule
		& 0.01 &0.02& 0.03& 0.04& 0.05&0.06&0.07&0.08&0.09&0.1 \\
		\midrule
		LVD \cite{li2021local}&0.704	&0.683	&0.665	&0.672	&0.667	&0.668	&0.678	&0.688	&0.676	&0.672 \\
		\midrule
		LVID &\textbf{0.841}	&\textbf{0.817}	&\textbf{0.804}	&\textbf{0.808}	&\textbf{0.804}	&\textbf{0.806}	&\textbf{0.815}	&\textbf{0.827}	&\textbf{0.821}	&\textbf{0.812} \\
		\bottomrule
	\end{tabular}
\end{table*}

\begin{table*}[]
	\centering
	\caption{The Kendell correlation coefficient when $\beta$ changes in Infectious network}
	\begin{tabular}{ccccccccccc}
		\toprule
		& 0.01 &0.02& 0.03& 0.04& 0.05&0.06&0.07&0.08&0.09&0.1 \\
		\midrule
		LVD \cite{li2021local}&0.809	&0.766	&0.746	&0.745	&0.732	&0.729	&0.726	&0.727	&0.720	&0.720 \\
		\midrule
		LVID &\textbf{0.852}	&\textbf{0.860}	&\textbf{0.859}	&\textbf{0.852}	&\textbf{0.838}	&\textbf{0.839}	&\textbf{0.829}	&\textbf{0.831}	&\textbf{0.827}	&\textbf{0.824} \\
		\bottomrule
	\end{tabular}
\end{table*}

\begin{table*}[]
	\centering
	\caption{The Kendell correlation coefficient when $\beta$ changes in PDZbase network}
	\begin{tabular}{ccccccccccc}
		\toprule
		& 0.01 &0.02& 0.03& 0.04& 0.05&0.06&0.07&0.08&0.09&0.1 \\
		\midrule
		LVD \cite{li2021local}&0.036	&0.055	&0.019	&0.035	&0.022	&0	&-0.020	&-0.023	&-0.026	&-0.045\\
		\midrule
		LVID &\textbf{0.544}	&\textbf{0.644}	&\textbf{0.689}	&\textbf{0.715}	&\textbf{0.717}	&\textbf{0.714}	&\textbf{0.707}	&\textbf{0.697}	&\textbf{0.683}	&\textbf{0.670} \\
		\bottomrule
	\end{tabular}
\end{table*}

\begin{table*}[]
	\centering
	\caption{The Kendell correlation coefficient when $\gamma$ changes in NS network}
	\begin{tabular}{cccccccccccc}
		\toprule
		& 0. &0.1& 0.2& 0.3& 0.4&0.5&0.6&0.7&0.8&0.9&1 \\
		\midrule
		LVD \cite{li2021local} &\textbf{0.727}	&0.731	&0.724	&0.730	&0.721	&0.711	&0.699	&0.717	&0.689	&0.679	&\textbf{0.672} \\
		\midrule
		LVID  &0.714	&\textbf{0.761}	&\textbf{0.767}	&\textbf{0.777}	&\textbf{0.774}	&\textbf{0.751}	&\textbf{0.752}	&\textbf{0.728}	&\textbf{0.713}	&\textbf{0.696}	&0.650\\
		\bottomrule
	\end{tabular}
\end{table*}

\begin{table*}[]
	\centering
	\caption{The Kendell correlation coefficient when $\gamma$ changes in Celegans network}
	\begin{tabular}{cccccccccccc}
		\toprule
		& 0. &0.1& 0.2& 0.3& 0.4&0.5&0.6&0.7&0.8&0.9&1 \\
		\midrule
		LVD \cite{li2021local} &0.666	&0.637	&0.617	&0.607	&0.594	&0.586	&0.589	&0.594	&0.579	&0.585	&0.570 \\
		\midrule
		LVID& \textbf{0.804}	&\textbf{0.761}	&\textbf{0.750}	&\textbf{0.734}	&\textbf{0.710}	&\textbf{0.696}	&\textbf{0.693}	&\textbf{0.701}	&\textbf{0.699}	&\textbf{0.700}	&\textbf{0.686}\\
		\bottomrule
	\end{tabular}
\end{table*}

\begin{table*}[htbp]
	\centering
	\caption{The Kendell correlation coefficient when $\gamma$ changes in Infectious network}
	\begin{tabular}{cccccccccccc}
		\toprule
		& 0. &0.1& 0.2& 0.3& 0.4&0.5&0.6&0.7&0.8&0.9&1 \\
		\midrule
		LVD \cite{li2021local} &0.733	&0.721	&0.707	&0.709	&0.710	&0.708	&0.722	&0.738	&0.751	&0.753	&0.747 \\
		\midrule
		LVID& \textbf{0.837}	&\textbf{0.820}	&\textbf{0.794}	&\textbf{0.784}	&\textbf{0.782}	&\textbf{0.764}	&\textbf{0.768}	&\textbf{0.770}	&\textbf{0.784}	&\textbf{0.774}	&\textbf{0.767}\\
		\bottomrule
	\end{tabular}
\end{table*}

\begin{table*}[htbp]
	\centering
	\caption{The Kendell correlation coefficient when $\gamma$ changes in PDZbase network}
	\begin{tabular}{cccccccccccc}
		\toprule
		& 0. &0.1& 0.2& 0.3& 0.4&0.5&0.6&0.7&0.8&0.9&1 \\
		\midrule
		LVD \cite{li2021local} &0.025	&0.014&0.028	&0.033		&0.022	&0.022	&0.028	&0.018	&0.021	&-0.004 &-0.020 \\
		\midrule
		LVID& \textbf{0.361}	&\textbf{0.358}	&\textbf{0.362}	&\textbf{0.365}	&\textbf{0.342}	&\textbf{0.357}	&\textbf{0.361}	&\textbf{0.346}	&\textbf{0.319}	&\textbf{0.283}	&\textbf{0.263}\\
		\bottomrule
	\end{tabular}
\end{table*}

\section{Conclusion}
In this paper, a novel approach for node importance evaluation in complex network, namely local volume information dimension, is proposed. The proposed method use information entropy to measure the information in the certain distance from the central node and considers information from different distance, which is superior to other methods. The superiority of the local volume dimension is validated by experiments on five networks obtained from the real-world. Encouraging results indicates the effectiveness of the proposed method with SIR model.

In the future, the proposed approach can be extended to more kinds of networks like signed networks, multiplex networks, temporal networks, hypergraphs and so forth. The estimation methods for the local volume information dimension can also explored to make LVID faster. Practical applications of LVID in community detection, link prediction, decision making and so forth, can also be explored. More types of entropy measures may also be used to judge node importance.

\section*{Acknowledgment}
The work is partially supported by National Natural Science
Foundation of China (Grant No. 61973332), JSPS Invitational Fellowships for Research in Japan (Short-term).
\section*{References}
\bibliographystyle{elsarticle-num}
\bibliography{References}
\end{document}